\documentstyle[aps,prl,twocolumn]{revtex}

\def\bfHcal{{\mbox{\boldmath${\cal H}$}}}
\def\be{\begin{equation}}
\def\ee{\end{equation}}
\begin{document}
\narrowtext
\draft 
\title{
Finsler Geometric {\it Local} Indicator of Chaos\\
for single orbits in the H\'enon--Heiles hamiltonian.
}
\author{P. Cipriani$^1$\cite{infm_rm2} and M. Di Bari$^2$\cite{infm_pr} }
\address{
$^1$Dipartimento di Fisica, Universit\`a "Tor Vergata" - 
Via della Ricerca Scientifica, 1\ -- \ {\bf 00133\ {\rm --}\ ROMA} \\
$^2$Dipartimento di Fisica, Universit\`a di Parma - 
Viale delle Scienze \ -- \ {\bf 43100\ {\rm --}\ PARMA}\\
}
\date{June 15, 1998}
\maketitle
\begin{abstract}
Translating the dynamics of the H\'enon--Heiles ({\sf HH}) hamiltonian
as a geodesic flow on a {\sl Finsler manifold}, 
we obtain a {\it local and synthetic} Geometric Indicator of Chaos 
({\sf GIC}) for two degrees of freedom ({\sf dof}) continuous dynamical systems 
({\sf DS}'s). It represents a link between {\sl local} quantities
 and {\sl asymptotic behaviour} of orbits
and gives a strikingly evident, {\sl one-to-one}, 
correspondence between geometry and instability. 
Going beyond the results attained using the customary dynamical approach 
and improving also on the {\sl global} criteria established within 
the Riemannian framework,  the {\sf GIC} is able to discriminate 
between regular and stochastic orbits on a given energy surface,
simply on the basis of the value it assumes along a relatively small piece of the trajectory,  
{\sl without {\it long} integrations of the dynamics} and without 
{\sl any reference to a perturbation}.
\end{abstract}
\pacs{{\tt PACS:}\ 05.45.+b\ ;\ 02.40.-k\ ;\ 02.90.+p}
\narrowtext

The presence of instability in {\sf DS}'s is usually 
recognized {\it a posteriori}, looking at the long time evolution of
{\sl small} perturbations, whose average
exponential growth is interpreted as {\it the} signature of Chaos.
Skipping here most of the issues related to the universal meaning
attributed to it, we remark how the procedures and tools having at the grounds
such a criterion alone manifest their limitations whenever is investigated the 
behaviour of {\sf DS}'s at the {\sl boundary} between {\sl quasi-integrability} and 
{\sl stochasticity} (e.g.,\cite{IndCaos,Aquila2}).
Instead, the search for {\it a priori, synthetic} signatures of Chaos, dates back to Toda
\cite{Toda-criticaToda} and continued, across interesting
investigations on the mechanisms of transition to stochasticity 
(see {\tt e.g.} \cite{UdryMartinet}), up to the recently revived
Riemannian {\sl Geometrodynamical approach} ({\sf GDA}) \cite{Marco_vecchi,Noi_vecchi}. 
Though most of its results relate to high dimensional 
hamiltonian systems (for which some approximations are justified by the large number 
of {\sf dof} or some {\sl weak form} of the ergodic hypotesis), more recently, 
this approach has been tested also for {\sl small} {\sf DS}'s, 
giving outcomes clearly supporting its reliability, 
\cite{Marco_2gdl,Aquila2}. Nevertheless, if in the
case of large {\sf DS}'s considered the agreement between the {\sf GDA} and the {\sl customary}
tools used to detect Chaos has revealed to be rather satisfactory, at least as long as 
the approximations are well justified
\cite{Aquila1}, some discrepancies emerge in the case of few 
{\sf dof} systems \cite{Aquila2,HH_CPMT}. 
For the latter, 
the {\sf GDA} provided an alternative way to recover {\sl most} of 
the results obtained with the tangent dynamics equations, suggesting deeper hints for
the understanding of the sources of Chaos and giving in addition some 
{\sl global} criteria to single out a transition in the overall behaviour \cite{Marco_2gdl}. 
However, this criterion is unable to correctly detect the occurrence of Chaos in single 
orbits\cite{nota_J_isole}, as it renounces, {\sl in principle}, to {\sl intrinsically}
describe the behaviour of {\sl individual trajectories}. 
Within the Riemannian approach to few {\sf dof} systems, 
this issue has been addressed, up to now, only resorting to a 
numerical procedure analogous to the integration of the tangent dynamics equations, 
whose results have been {\sl generally} confirmed (though not always). 
Recently Kandrup \cite{Kandrup97} investigated in details the
 relationships existing between {\it local} {\sl dynamical behaviour}
and {\it local} {\sl geometric features} of the Jacobi (Riemannian) manifold for some
 two-dimensional {\sf DS}'s, obtaining 
qualitative correlations among average curvature and its fluctuations
and somewhat more ambiguous ones between curvature fluctuations and {\sl short time
Lyapunov exponents}.
In summary,  the Riemannian {\sf GDA} has been able, up to now, either to {\sl intrinsically}
describe the {\sl average behaviour} of a {\sf DS} or to single out the {\sl individual}
orbits instability {\it a posteriori}, as in the Hamiltonian description.
Both these approaches have revealed unable to find an {\sl intrinsic and individual} 
indicator of {\sl long-term behaviour of orbits} as the {\sf GIC} here presented, 
built within the Finsler {\sf GDA} and 
which {\sl cannot even to be defined}, (for the {\sf HH} system) within the
other above mentioned frameworks.  
Notwithstanding its {\sl local} character (in both spatial and temporal meanings), 
it reveals to be unambiguously related to Lyapunov Characteristic Numbers ({\sf LCN}'s), 
{\it i.e.}, to asymptotic quantities, usually computed with reference to a perturbation! 
We claim then that it represents a strong indication (if not a proof)
 that the {\sf GDA} is able not only to {\sl reproduce} and to {\sl explain}
the results obtained with the usual tools, but even to go beyond them.

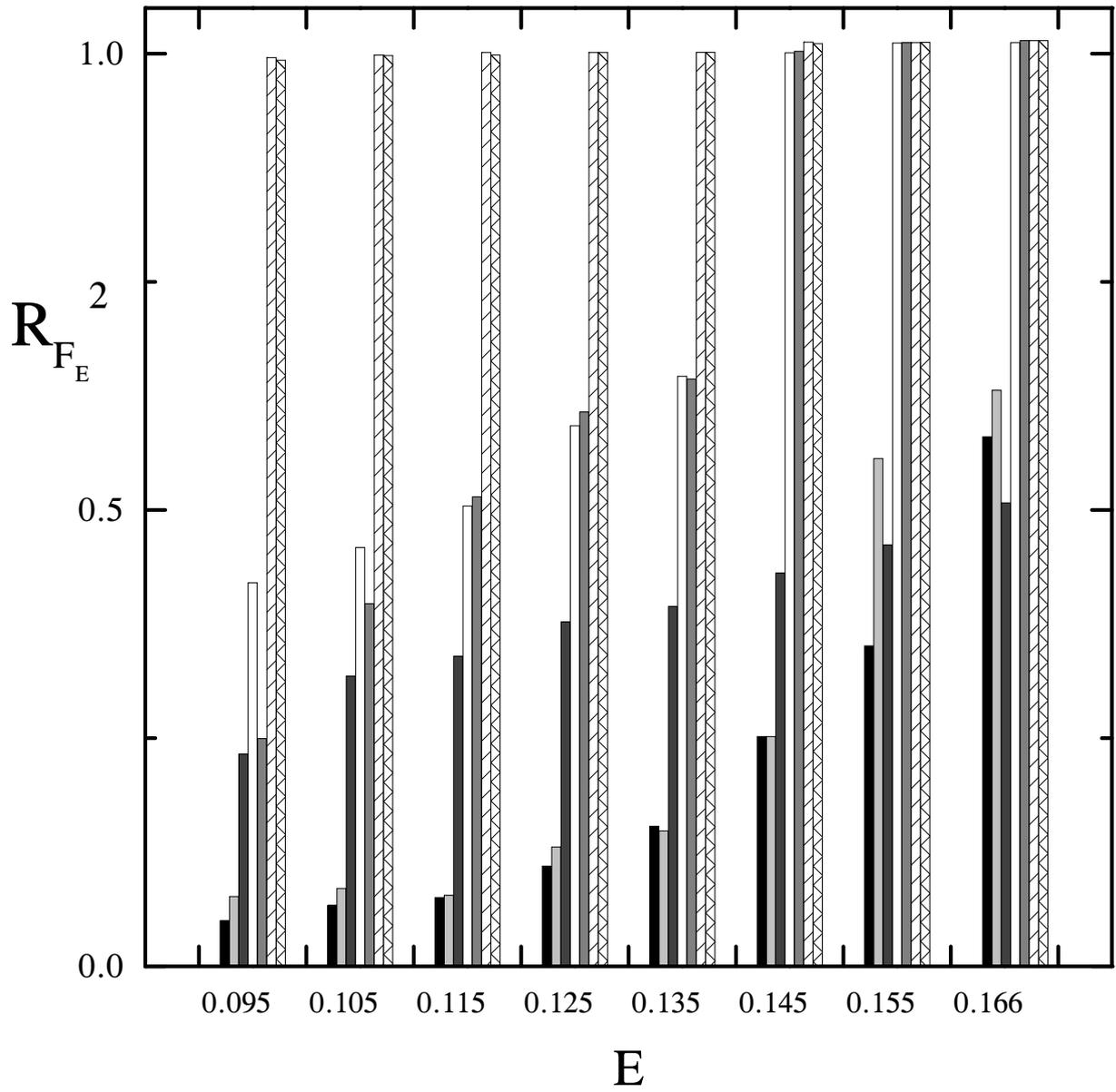
\begin{figure}
\caption{
Values of $R_{F_E}^2$ for geodesics starting from i.c. "topologically equivalent" to those
indicated on figure 1 for $E=1/8$, at energies from $E=0.095$ up to $E=0.166$.}
\label{fig:RFdiE}
\end{figure}

One of the main recent results of the {\sf GDA} is the confirmation that
the onset of unpredictability in the geometric transcription of realistic 
(large) {\sf DS}'s is driven by the mechanism of parametric instability
\cite{Marco_vecchi,Noi_vecchi} and thus differs completely from what occurs
in the geodesic flows of abstract Ergodic theory.
Indeed, most of phase space of large {\sl physical} {\sf DS}'s is not characterized 
by (constant) negative curvatures, but stochasticity is caused by the {\sl quasi random}
fluctuations of (mostly) positive curvatures. 
For few dimensional {\sf DS}'s however, such random character cannot be assumed and 
instead is {\sl parametric resonance}, similar to that occurring in the Mathieu 
equation, to bring about instability, \cite{Noi_vecchi,Marco_2gdl,PRE97}. 
Nevertheless, we find below that even for two {\sf dof} systems such a mechanism cannot
be singled out in a {\sl na\"\i ve} way and very intricate combinations
of geometric features of the manifold are linked to instability in a non trivial
way.\\
 We already discussed the motivations for an extension of the {\sf GDA} to 
include non Riemannian manifolds \cite{PRE97} and we also pointed out its 
somewhat {\sl greater effectiveness} with respect to the {\sl usual} tools in the 
computations of {\sl instability exponents} \cite{Aquila2,BIX_PRL}. 
We refer to \cite{Marco_vecchi,Noi_vecchi,Aquila1} 
for a detailed description of the {\sf GDA}, to \cite{Aquila2,PRE97,MTTD} 
for its implementation within Finsler manifolds
and to \cite{HH_CPMT} for a thorough discussion of most of the points
here only sketched.\\ Within the {\sf GDA} the trajectories of a ${\cal N}$ {\sf dof} system
become the geodesics of suitable differential manifolds, which, in Finsler geometry,
are $({\cal N}+1)-$dimensional and represent a generalization of Riemannian ones. 
The stability of the flow is determined
by the {\sl Jacobi--Levi-Civita} ({\sf JLC}) equations of geodesic spread:
\be
{\nabla\over{ds}} \left( {{\nabla z^a}\over{ds}} \right)~ +~
 {\cal H}^a{}_c ~z^c~ =~ 0\ ,\ \ \ (a=0,1,\ldots,{\cal N})\ ,
\ee
being $z^a$ the perturbation, $\nabla/ds$ the covariant
derivative {\sl along the geodesic} and the {\sl stability tensor},
\bfHcal, \cite{Noi_vecchi,Aquila1}, derives from the (generalized)
curvature tensor of the manifold. 
The Finsler {\sl time-parameter} $s\equiv s_F$ is defined through the Lagrangian function
${\cal L}$, as $ds_F = {\cal L}\,dt $ and
possesses a built-in invariance with respect to an arbitrary rescaling of the
{\sl Newtonian} time $t$, \cite{nota_rescaling}. 
The {\sl local} behaviour of geodesics
is determined by the eigenvalues of \bfHcal, which are the 
{\sl principal sectional curvatures} ({\sf psc}'s) defined {\it by the given geodesic}
on the manifold \cite{Aquila1}.
For a ${\cal N}$ degrees of freedom {\sf DS}, once a geodesic is chosen, the Finsler
stability tensor possesses $({\cal N}+1)$ eigenvalues, $(\{\lambda_A\},\ A=0,1,\ldots ,{\cal N})$
one of which vanishes identically, $\lambda_0\equiv 0$, associated with a
{\sl neutral} eigenvector,
along the tangent to the geodesic. In the case of a
{\sl standard} hamiltonian system ({\it i.e.} without gyroscopic terms),
$ H({\bf q},{\bf p}) = {1\over {2}}~ a_{ij} p^i p^j + {\cal U}({\bf q})$
we have then, \cite{Aquila2,PRE97}:
\begin{equation}
\lambda_i = t' (B + t' \mu_i)\ ,\ \ (i=1,\ldots,{\cal N})
\end{equation}
where $B$ is related to the time derivatives of
the Lagrangian, the $\{\mu_i\}$ are the eigenvalues of the hessian 
${\cal U},_{ij}({\bf q})$ and the prime denotes the derivative wrt to $s_F$.\\
In this letter we will deal with the well known two dimensional {\sf HH} system, 
whose Hamiltonian is
\be
H({\bf q},{\bf p}) = {1\over{2}} \left( x^2 + y^2 + p_x^2 + p_y^2
\right) +x^2y - {{y^3}\over{3}}.
\ee
We find that for this {\sf DS} (as well as for generic realistic ones, 
in spite of some persistent claims, e.g. \cite{Junqing97})
negative curvatures are quite unable to explain the asymptotic character of orbits. 
The mechanism of {\sl parametric instability}, due to fluctuations of usually
positive curvatures \cite{Marco_vecchi,Noi_vecchi}
manifests {\it na\"\i vely} however only for some
many {\sf dof} systems, being instead hardly perceived
 in this case. For example, the analysis of spectra, \cite{HH_CPMT}, 
shows how intermingled and far from trivial are the relationships with the 
{\sl elementary} theory of {\sl Mathieu-like} equations.
Such seemingly discouraging inconsistencies driven us
to check for different signatures of instability. 
The Finslerian approach allows to consider, even for two {\sf dof} Hamiltonians,
the possible anisotropy of the manifold, which results to
play a crucial role, via the {\sl Schur theorem} \cite{Aquila1}
in the mechanism of instability:  fluctuating curvatures require also
that the manifold is anisotropic. The connection between curvatures variations along
a geodesic and anisotropy, on one side, and growth and {\sl rotation} of perturbation, 
on the other,
is far from being clear and is currently investigated: how the former can interact
to steer the latter can at the moment only be guessed. For a two {\sf dof} system, the 
associated Finsler manifold is three dimensional and its curvature properties
{\sl along a {\it given} geodesic} are described by the two non vanishing 
{\sf psc}'s, $\lambda_{1,2}$, which are invariant functions on the tangent space, 
representing the sectional ({\it i.e.} gaussian) curvatures in the two-planes defined 
from the flow and the two (non trivial) eigenvectors of \bfHcal. Given them, we can
characterize the way the geodesic explores the manifold
 through the (half) {\sl Ricci curvature} (along the flow) and the {\sl anisotropy}, 
$\kappa[{\bf q}(s), {\bf p}(s)]$ and $\vartheta[{\bf q}(s), {\bf p}(s)]$, respectively:
\be
\kappa\doteq {{\lambda_1+\lambda_2}\over {2}} \equiv {{{\rm Tr}(\bfHcal)}\over {2}}
= {{{\sf Ric}_{\rm F}({\bf u})}\over {2}}
\ \ \ ;\ \ \  
\vartheta\doteq {{\lambda_1-\lambda_2}\over {2}}\ .
\ee
An exhaustive statistical analysis of the behaviour of $\kappa$ and $\vartheta$ (or
equivalently of $\lambda_{1,2}$) and the details of the logical path leading
to the {\sl synthetic indicator} are presented elsewhere, \cite{HH_CPMT}.
We found that along a generic geodesic the $\{\lambda_i\}$ oscillate around their
average values, the fluctuations of Ricci curvature in general turns out to be however
much smaller than those of the sectional ones, which are indeed almost anti-correlated.
Such an effect is particularly evident in the {\sf HH} case, as 
$\Delta{\cal U}\equiv 2$. So, the manifold appear to be everywhere anisotropic but
with {\sf psc}'s always ($\lambda_1$) or mostly ($\lambda_2$) positive.
Fluctuations  (and then anisotropy) increase with energy, showing a {\sl global}
qualitative change in correspondence of appearance of stochasticity, \cite{HH_CPMT}.
Although smaller, the overall fluctuations of Ricci curvature seem nevertheless to
influence appreciably the stability of geodesics. Moreover, Schur theorem asserts that
the two quantities must be related. We then look for a relative measure of
anisotropy fluctuations compared to overall curvature variations.
Correlations between two quantities $A(s)$ and $B(s)$ 
reflect in the functional:
\be
\widetilde {\cal C}_S [A,B] ~\doteq~ 
{{\langle A\cdot B\rangle_S}\over{\left(\langle A^2\rangle_S~\cdot~\langle 
B^2\rangle_S\right)^{1/2}} }\ ,
\ee
which clearly depends on the averaging interval $S$, dependence understood in the sequel. 
An indication about the relative importance 
of anisotropy wrt average curvature fluctuations can be obtained comparing the 
phase space normalized correlation functions  
$\widetilde {\cal C}[\vartheta, \delta\vartheta]$
and  
$\widetilde {\cal C}[\kappa, \delta\kappa]$,
being $\delta A(s)~\doteq~ A(s) - \langle A\rangle$. 
Using them we build up a quantity which apparently
describes only the {\sl local} geometric features of the sub-manifold explored  
(in a lapse $S$)
by the geodesic having initial conditions 
$[{\bf q}(0),{\bf p}(0)]=({\bf q}_{{}_0},{\bf p}_{{}_0})$:
\be
R_F[S]~\equiv~R_F({\bf q}_{{}_0},{\bf p}_{{}_0})~\doteq~{
{\widetilde {\cal C}[\vartheta, \delta\vartheta]}
\over
{\widetilde {\cal C}[\kappa, \delta\kappa]}
}~\geq~0
\ee
However, the inspection of figure \ref{fig:E=1/8} shows a striking correspondence 
between the {\sf PSS}'s and the map of $R_F^2$.
Indeed, figures \ref{fig:E=1/8} a) and b), which refer to the
{\sl typical} energy value $E=1/8$,  show that smaller is the value of 
$R_F({\bf q}_{{}_0},{\bf p}_{{}_0})$,
{\it more regular} is the geodesic passing through $({\bf q}_{{}_0},{\bf p}_{{}_0})$.
To obtain the plot of figure \ref{fig:E=1/8} b), we chosen a grid of points 
$\{y_{{}_0},p_{y_0}\}$ on the {\sf PSS} $x=0$, chosing $p_{x_0}$ such that $H=E$ and
then numerically integrated the geodesic equations, computing the
correlation functions entering $R_F({\bf q}_{{}_0},{\bf p}_{{}_0})$. 
The same results have been obtained at all the energies; the more {\it essential}, though
less {\it suggestive}, histograms of figure \ref{fig:RFdiE}, confirm that our
{\sf GCI} is able to depict correctly the single orbits behaviour up to the
dissociation energy. 
The columns in this plot represent the values of $R_{F_E}^2$,
defined as 
\be
R_{F_E}^2({\bf q}_{{}_0},{\bf p}_{{}_0})~\doteq~ 
\alpha^2(E)~ R_F^2({\bf q}_{{}_0},{\bf p}_{{}_0})\ ,
\ee 
where $\alpha(E)=\alpha_1\cdot E^2$ is a scaling factor to get rids of a 
{\sl global} energy dependence, useful to compare results at different
energies.
For each energy, the $R_{F_E}^2$ values are reported for a sample of seven initial conditions,
 chosen among those {\sl topologically equivalent} to the ones indicated in figure 
\ref{fig:E=1/8}b) for $E=1/8$. We observe that while
chaotic trajectories are always characterized by $R_{F_E}^2$ values around its upper
limit (normalized to unity, this is because we use $R_{F_E}^2$ instead of $R_{F}$), regular
orbits have instead considerably smaller values, which is however higher for those geodesics
{\sl tending to become chaotic earlier}, as the energy increases.
In particular, from both the figures, we see that the geodesics 
in the {\sl regular islands} located on the
$p_y-$axis of the {\sf PSS} are recognized by our {\sf GIC} as {\sl nearer} to chaotic
ones than those belonging either to the large regular island on the $y-$axis or to the
{\sl banana} region; this explains why these islands {\sl disappear} earlier as the energy
increases and gives also some insights on the causes of the (partial) 
failure of the Riemannian approach to describe these {\sl peculiar} 
orbits \cite{Aquila2,Marco_2gdl,nota_J_isole}. 
Moreover, very
interesting insights can be obtained looking at the {\sl relaxation patterns}
of $R_F$ as function of the averaging time $S$ \cite{HH_CPMT}, as can be perceived from
the {\sl diffusive} behaviour around the border of regular islands.
Well then, a geometric {\sl local} quantity turns out to
be deeply related to the {\sl asymptotic} behaviour of geodesics.
We stress that the values plotted in the map and histograms are obtained through 
computations of correlation functions over time intervals much shorter than those 
needed to obtain either the {\sf PSS}'s or the {\sf LCN}'s: 
it suffices to follow a geodesic for as few as fifteen periods in order to 
see if the value of $R_F$ attains the upper limit characterising the {\sl stochastic sea} 
or tends to a lower value, which though different for distinct regular orbits, 
is however always smaller than for chaotic ones. A {\sl local signature}
 of asymptotic instability acquires a special significance also on the light of the
issue of reliability of long time numerical integrations of chaotic systems \cite{Sauer97}.
The definition of $R_F$ obviously implies that it can be used also as a {\sl global} indicator,
able to give a quantitative measure of the overall degree of stochasticity at a fixed
value of energy: a phase-averaged {\sl local} indicator is a {\sl global} one,
instead a global indicator is, in general, {\sl non-local}. 
Perplexities which can be raised by the apparently cumbersome definition of $R_F$ 
are answered on the light of the {\sl pathologies} affecting the {\sf HH} hamiltonian,
 which amount mainly to the degeneracies of its integrable limit \cite[p.46]{LiLi}. 
Indeed, for most two dimensional {\sf DS}'s, more {\sl na\"\i ve} Finsler geometric 
indicators suffice to discriminate between chaotic and regular orbits.
Although the outcomes here presented (and in \cite{Aquila2,HH_CPMT}) clearly support the
reliability of Finsler {\sf GDA}, nevertheless a better understanding is
still waiting, as we need to test and extend the proposed criterion to more
general and higher dimensional {\sf DS}'s. Moreover it is desirable to
improve the rather phenomenological interpretation of $R_F$ and possibly to
predict theoretically the {\sl threshold} above which stochasticity occurs. This goal
amounts essentially to understand whether high $R_F$ values either follow from or are 
a cause of stochastic behaviour (or both).
The results obtained support the conviction 
\cite{Marco_vecchi,Noi_vecchi,Marco_2gdl,Aquila1,Aquila2} 
that negative curvature is unnecessary (if not irrelevant) to explain 
the onset of Chaos in realistic {\sf DS}'s, keeping however in mind that some faded global
correlations exist. Instead, a separate spot concerns the long-lasting claims about the 
implications of scalar curvature: except for two dimensional or isotropic manifolds, 
it appears to have nothing to do with the behaviour of geodesics representing realistic 
{\sf DS}'s \cite{Aquila1} and an hopefully coherent explanation of its irrelevance will be
presented elsewhere.
Among the issues still open, it is worth to investigate more deeply the relevance of 
the rotation of eigendirections of \bfHcal: a direct inspection of the {\sf JLC} equations
seems to indicate that the rotation of the perturbation vector should give a 
contribution to instability, nevertheless a {\sl separation by hand} of 
{\it interacting effects} often causes serious inconsistencies.

\begin{figure}
\caption{
Phase space portraits of {\sf HH} system at $E=1/8$:\\
{\bf a)} {\sf PSS} obtained numerically integrating a sample of orbits up 
to {\it T}=10000 units;
{\bf b)} Gray-scale plot of $R_F^2(y_{{}_0},p_{y_0})$ computed following a
set of geodesics for a lapse {\it S}=200 units. 
Darker dots represent smaller $R_F$ values, the {\sl White Stochastic Sea} corresponds
to $R_{F_E}^2 > 0.8$. The numbers label the initial conditions of figure 2.
}
\label{fig:E=1/8}
\end{figure}

\begin{references}
\bibitem{nota_J_isole} One of the outstanding outcomes of Pettini and 
coworkers investigations
resides in the possibility of finding indicators of Chaos through
{\sl microcanonical averages}, without integrations of trajectories. It is an obvious
price to
pay for this achievement to renounce, {\sl in principle}, to study the local
behaviour of the {\sf DS}, looking only to its average global features.
The indicator found in \cite{Marco_2gdl} fails indeed to 
distinguish between regular and stochastic orbits on a given energy surface:
it assumes in chaotic regions of the phase space a value
in between those taken in two (disjoint) regular islands \cite{HH_CPMT}.  
This is confirmed in \cite{Aquila2} where a thorough exploration of the 
{\sf HH} phase space reveals somewhere quantitative (and even qualitative)
 disagreements between dynamical behaviour and 
Riemannian description.
\bibitem{nota_rescaling} The non trivial effects consequent to an
arbitrary rescaling of the time variable are well known within a general relativistic
context \cite{Aquila2,BIX_PRL} but should be considered also in Classical Mechanics.
\end{references}
\end{document}